\documentstyle[11pt]{article}
     
\def\slashchar#1{\setbox0=\hbox{$#1$}           
   \dimen0=\wd0                                 
   \setbox1=\hbox{/} \dimen1=\wd1               
   \ifdim\dimen0>\dimen1                        
      \rlap{\hbox to \dimen0{\hfil/\hfil}}      
      #1                                        
   \else                                        
      \rlap{\hbox to \dimen1{\hfil$#1$\hfil}}   
      /                                         
   \fi}
\begin{document}
\begin{titlepage}
\begin{center}
      \hfill     BUHEP-97-24\\
\vskip 0.2 in
{\Large \bf Towards an Effective Particle-String Resolution \\
of the  Cosmological Constant Problem}

\vskip .2 in
         {\bf   Raman Sundrum}
        \vskip 0.3 cm
       {{\it Department of Physics \\
Boston University \\
Boston, MA 02215, USA} \\
~\\
email: sundrum@budoe.bu.edu}
 \vskip 0.7 cm 

\begin{abstract}
The Cosmological Constant Problem is re-examined from an effective field
theory perspective. While the  connection between gravity and 
 particle physics has not been experimentally probed in the quantum
 regime, it is severely constrained by the successes of Standard Model
 quantum field theory at short distances,  and classical General
 Relativity at large distances. At first sight, it appears 
 that combining particle physics
 and gravity inevitably leads to an effective field theory below the
 weak scale which suffers from large radiative corrections to the 
 cosmological constant. Consequently, this parameter must be very 
finely tuned 
to lie within the experimental bounds. An analog of just this type of 
predicament, and its resolution, are described in some detail using only
familiar quantum field theory. The loop-hole abstracted from the analogy
is the possibility of graviton ``compositeness'' at a  scale less than
$10^{-2}$ eV, which cuts off the large contributions to the cosmological
constant from standard model physics. Experimentally, this would show up as
a dramatic breakdown of Newton's Law in upcoming sub-centimeter tests
of gravity. Currently, strings are the only known example of such
compositeness. It is proposed that the gravitational sector comprises
 strings of very low tension,   which couple to a stringy ``halo'' 
surrounding each point-like standard model particle.

\end{abstract}
\end{center}

\end{titlepage}

\section{Introduction}

A naturalness  problem is like the sight of a needle standing upright
 on a table; it is {\it consistent} to assume a delicate balance,
but one strongly suspects an invisible stabilizing force. The balance
one must explain has the form of an extremely fine cancellation between
large virtual contributions to an observable from physics at 
very different length scales. 
The grandest and most baffling of all the naturalness problems in
fundamental physics is the Cosmological Constant Problem (CCP). It emerges 
upon putting together the
two separately successful parts of our physical understanding: classical
General Relativity and the quantum field theoretic Standard Model (SM). 
Ref. \cite{review} provides a good review. 
The problem is so tightly constrained that one can hope
that its final 
resolution will reveal an essentially unique and qualitatively new 
 stabilizing mechanism. 

Here is an outline of the problem. The classical theory of  general
relativity that has been  tested at long distances can be thought of as
the result of integrating out short distance quantum fluctuations  from
both the SM and gravitational sectors.
Einstein's equations describing the leading long-distance
behavior of the metric field, $g_{\mu \nu}$, are, 
\begin{equation}
R_{\mu \nu} - \frac{1}{2} g_{\mu \nu} R =  \frac{1}{4 M_{Pl}^2}
[\lambda ~ g_{\mu \nu} + T_{\mu \nu}].
\end{equation}
Here $R_{\mu \nu}$ and $R$ are the curvature tensor and scalar
respectively, $T_{\mu \nu}$ is the classical 
energy-momentum tensor for the matter
and radiation in the universe, Newton's constant has been written in
terms of  the Planck mass, $M_{Pl}$, 
and $\lambda$ is the cosmological
constant.\footnote{This definition of the cosmological constant differs
  by a factor of $4 M_{Pl}^2$ from the astrophysical convention, in
  order to give $\lambda$ the units of  energy-density.}
 $\lambda$ is very
sensitive to the short distance physics which has been integrated
out. The SM contributes its ``vacuum
energy'', very roughly given by,
\begin{equation}
\lambda_{SM} \sim {\cal O}(v^4),
\end{equation}
where $v$ is the scale of electroweak symmetry breaking. The other
contributions are from less well-understood sources, namely  short-distance
quantum gravity and particle physics beyond the standard
model.\footnote{My language assumes that the standard model (possibly
  with the exclusion of the physical Higgs degree of freedom) is just an
  effective theory valid below roughly $v$, and is superseded by some
  other theory at higher energies.} However, as long as these exotic
contributions do not 
unexpectedly  finely cancel  the SM contribution, we
must have the following rough lower bound on the cosmological constant, 
\begin{equation}
|\lambda| \geq {\cal O}(v^4).
\end{equation}
On the other hand, in solutions to eq. (1), 
$\lambda$ contributes to the cosmological expansion
rate. This permits a conservative bound to be put on $\lambda$ by
 using the
measured expansion rate of the universe and estimates of $T_{\mu \nu}$.
 With high confidence, the experimental bound is given by
\begin{equation}
|\lambda| < 10^{-56} v^4. 
\end{equation}
Now, eqs. (3) and (4) are in wild disagreement. To lower the bound in
eq. (3) sufficiently to accord with eq. (4) requires an unbelievably
fine cancellation between the 
contributions to $\lambda$ from quantum fluctuations below and above the
weak scale. 
This is  the CCP. 

The  attitude taken in this paper is that some part of the preceding
story is simply wrong, and the true story must  eliminate
 the need for fine-tuning in order to obtain an acceptably small value
 for $\lambda$. It is frequently believed that the true account cannot
 be understood by conventional means. According to this view, the
 resolution of the CCP may not be expressible in terms of the local
 quanta and interactions of a relativistic quantum theory. This is not
 the viewpoint of the present paper; the fundamental principles of
 relativity, quantum mechanics and locality are central to the
 understanding of the CCP, and the proposed resolution does not
 transcend them.

However, the CCP as described above is extremely robust, based only on
the co-existence of gravity and mass scales of order $v$ (and
supersymmetry-breaking of  at least the same magnitude). The CCP then 
follows by
elementary power-counting. Section 2 of this paper describes the CCP in
greater detail using effective field theory methodology. Effective field
theory provides a clear and economical separation of the facts and
principles which we have already tested experimentally, from the physics
which is still beyond our reach, both in the gravitational and particle
physics  sectors. It is a useful language for examining assumptions
which we may need to discard, as well as for evaluating new proposals
for solving the CCP. 

In order to solve the CCP we must change the power-counting which
determines how sensitive to SM mass scales the long-distance theory of 
eq. (1) is. In quantum field theory, whenever the physics of a large
mass scale is integrated out, the sensitivity of the low-energy
effective theory is determined by power-counting for the weakly-coupled
degrees of freedom {\it at that mass scale}, not just 
the degrees of freedom at the lowest energies. To use this observation
in the case of the CCP, we must ensure that at particle physics
energies, the gravitational degrees of
freedom are profoundly different from those in eq. (1). We can loosely
speak of the graviton as a ``composite'' of these new degrees of
freedom. How can this crucial new physics be right underfoot without our
having noticed, and how exactly can compositeness help with the CCP? 
Section 3
provides a detailed analogy of the CCP where these questions, and
others, 
can be understood in the context of a simple toy universe. 
This serves as a useful warm-up because 
the resolution of the toy naturalness problem is based on
completely familiar physics. 

Finally, in section 4, a possible new mechanism  
for stabilizing an acceptably
 small cosmological constant is put forward. Below the weak scale, it
 consists of a gravitational sector made of extremely low-tension
 strings, with   string-scale less
 than $10^{-2}$ eV, interacting with  the stringy ``halos'' carried by SM
 particles. The low string tension  cuts off the virtual
 contributions to $\lambda$ so that it naturally satisfies eq. (4). 
At very large distances only the massless
 string mode, namely the graviton, is relevant, and the dynamics reduces
 to general relativity. In accelerator experiments, the macroscopic
 string halo carried by particles is unobserved because the stringy
 gravitational 
 physics is too weakly coupled to compete with point-like SM interactions.
The detailed structure of such an effective particle-string theory has
not yet been worked out, but I discuss its necessary properties as well
as possible directions towards its construction. If this scenario is
correct there will be a striking experimental signature: Newton's Law 
will completely break down when
gravity is tested at sub-centimeter distances!

This proposal may appear heretical from the view of traditional field
theory and string theory.
However, recent  developments in
string theory offer some encouragement. There is evidence that strings
can co-exist with objects of different dimensionality, D-branes,
including $0$-branes which are point-like states. For a review see
ref. \cite{dbranes}.
There have already
been several calculations of the scattering of 
D-branes with strings and with each other,  
which reveal a stringy halo about the D-branes
\cite{douglas} \cite{scattering}. These results may be useful
for constructing effective particle-string theories. Refs. \cite{ps} are
some initial forays in this direction. 

 However, I wish to point
out an important difference between the goal of the present work
and the goal of most of the string literature.
 The
recent string theory upheaval is part of a very ambitious program
aimed at a non-perturbative understanding of fundamental interactions at
at the highest energies. 
On the other hand, the CCP is a puzzle whose answer lies at
present-day energies, but is presumably 
hidden from view because of the weakness of the gravitational force. 
 The purpose here is to
develop an {\it effective} theory which has an ultraviolet cutoff given
by the weak scale, and whose parameters can {\it naturally} be fit to
experiment. The effective theory is permitted to break down above the
weak scale, and be replaced by a more fundamental description there. 

 Sections 2, 3 and 4 may be read in any order
depending on the background and interests of the reader.  Section
5 provides the conclusions.

I make  use of  rough estimates in several places.
It is customary when power-counting to keep track of factors of $4 \pi$
that arise from the dimensionality of space-time. In this paper, I will
consider all such factors to be order one since the Cosmological Constant
Problem involves such large numbers that, by comparsion, $4 \pi$ factors
are unremarkable. When estimating Feynman diagrams, dimensional
regularization is implicit for simplicity. This will  not remove any
important physics (for example it does not eliminate the CCP) because
the important mass scales  will be explict
and will not need to be represented by a dimensionful cutoff. 

\section{The Problem in Context}

\subsection{The standard effective theory of particles and gravity}

The most straightforward way to put together the SM and
general relativity is to write the lagrangian
\begin{equation}
{\cal L}_{eff}(v) = \sqrt{-g} \{\lambda_0 + 2 M_{Pl}^2 R + 
{\cal L}_{SM} +
...\},
\end{equation}
where $g_{\mu \nu}$ appears in ${\cal L}_{SM}$, minimally coupled to
maintain general covariance.\footnote{To be more precise, for fermions
  we must work in terms of the vierbein, but this detail is inessential
  in this paper.} In
order to compute quantum mechanical fluctuations of the metric around a
Minkowski space vacuum we note that,
\begin{equation}
g_{\mu \nu} = \eta_{\mu \nu} + \frac{h_{\mu \nu}}{M_{Pl}}, 
\end{equation} 
where $\eta_{\mu \nu}$ is the Minkowski metric, and $h_{\mu \nu}$ is the
canonically normalized spin-$2$ graviton field. For most regions of
spacetime, this weak field expansion about a Minkowski metric is justified.
The broad perspective of general relativity adopted in this paper, as a
phenomenological theory of gravity, is detailed in
refs. \cite{weinberg} \cite{veltman} \cite{donoghue}. It is based on the
generally observed principles of relativity and quantum mechanical
unitarity.

As is well-known, the inclusion of gravity renders the lagrangian
non-renormalizeable by elementary power-counting. 
This means that the resultant theory
cannot be a fundamental description of nature at all energies (at least
perturbatively). However, the lagrangian is a sensible basis for a
quantum theory effective at energies far below the Planck scale,
$M_{Pl}$. Recall how this
works in a general non-renormalizeable theory.
Technically a non-renormalizeable theory requires an infinite number of
counterterms, which at first sight appears disasterous. The situation
greatly improves if we restrict ourselves to physical 
processes at energies, $E$,
far below the (smallest) mass scale suppressing the non-renormalizeable
interactions, $M$. This allows us to work to any fixed order in the small
parameter $E/M$, say ${\cal O}((\frac{E}{M})^n)$. To this order only the
{\it finite} number of interactions and counterterms of dimension less than or
equal to $n + 4$ are relevant. While this statement is rather obvious at
tree-level, non-trivially it survives loops and renormalization. For $E
\ll M$ we thereby obtain a well-defined and predictive {\it effective}
  field theory. The effective theory must give way to a more
fundamental description at some scale below $M$, or perhaps be sensible
but strongly-coupled above $M$. The best known example of a 
non-renormalizeable effective field theory
is the chiral lagrangian description of pions, treated as Nambu-Goldstone
bosons of chiral symmetry breaking. For a review see
ref. \cite{hpet}. 
The typical scale appearing in the
non-renormalizeable interactions is the hadronic scale, $M \sim 
1$ GeV. 
 The effective field theory is therefore sensible and
weakly-coupled for $E \ll 1$ GeV. For $E > 1$ GeV, the
effective theory fails completely and must be replaced by the more
fundamental QCD description.

In the case at hand, the scale suppressing the non-renormalizeable
interactions is $M_{Pl}$. Therefore the theory given by eq. (5) 
makes sense at
energies $E \ll M_{Pl}$ \cite{donoghue}.
In fact let us take the ultraviolet cutoff of
our effective theory to be the weak scale, as denoted by
the $v$ appearing on the left-hand side of eq. (5). This allows us to
remain agnostic about the nature of physics beyond the weak scale. The
ellipsis in eq. (5) can contain
higher-dimension gauge and coordinate invariant interactions, whose
effects are small  at energies far below the weak scale.\footnote{More
  precisely, they are {\it irrelevant} in that their dominant effects
  can be absorbed into finite renormalizations of the lower dimension
  interactions.}

As far as accelerator experiments are concerned, eq. (5), provides a
very economical summary of what has been actually
observed. They overwhelmingly confirm  a relativistic  quantum field
theory given by the SM for energies below the weak scale,
with gravitational forces being negligible. Thus the ``laboratory
tested'' part of eqs. (5, 6) is given  by the formal limit, 
\begin{equation}
{\cal L}_{eff}^{lab}(v) \rightarrow  {\cal  L}_{SM}, 
~~~ g_{\mu \nu} \rightarrow \eta_{\mu
  \nu}, ~~~ {\rm as} ~~ 1/M_{Pl} \rightarrow 0.
\end{equation}

At macroscopic distances, with large amounts of matter and radiation, 
the SM forces are
effectively neutralized and gravity dominates. Because of the large
distances, masses and numbers of quanta, the classical approximation is
justified. Conceptually, one arrives at a {\it classical} effective
theory for this regime by integrating out all quantum fluctuations from
eq. (5). The result must have the form of classical general relativity,
eq. (1). This is because eq. (1) is the most general form consistent
with the general covariance of our starting point, eq. (5), up to terms
involving higher-dimension metric invariants which
 are irrelevant at macroscopic distances. 

It is important to note that only the classical macroscopic 
effective theory, rather
than the full effective quantum field theory of eq. (5), has been tested
gravitationally. This is in contrast to the SM sector, where the full
quantum field theoretic implications of eq. (5) have been
tested. Therefore we must bear in mind that while eq. (5) is in accord
with all gravitational tests since it reduces to eq. (1), the
``bare'' parameter $\lambda_0$ allowing us to fit the experimental bound of
eq. (4), eq. (5) may not be unique in this respect.

It is somewhat of a nuisance that eq. (1) combines two steps in its
derivation, the integrating out of microscopic
physics and the classical limit for large numbers of quanta. It is
useful to separate the two issues by considering a long distance
 effective theory
in a simplified setting, involving just a few SM particles,  but treated
fully quantum field theoretically. I will develop such an effective
theory in the next subsection. It will provide a useful point of contact
when we discuss the analogy in Section 3.

\subsection{A macroscopic quantum effective lagrangian with gravity}

Consider a few stable massive  spin-$0$ particles, $H$, with {\it relative}
momenta only of order $\mu \ll v$, interacting with soft gravitons with
energies of order $\mu$. The length scale, $1/ \mu$ will act as our
short-distance cutoff. We can take $1/\mu \sim 1$ mm, which
is less than the shortest range over which gravity has been tested. 
 One can imagine
$H$ to be a ground state hydrogen atom say, whose compositeness cannot
be resolved by the long wavelength gravitons. 
Alternatively we can take $H$ to be just a proxy for
more fundamental particles like an electron, neglecting the complications
of spin and charge. For simplicity I will also neglect the other soft
massless particles, photons and neutrinos. We therefore have an
isolated sector  which  should be described by
an effective lagrangian containing only the $H$ and $g_{\mu \nu}$
fields. This
type of theory is entirely analogous to the heavy particle effective
theories used in studying the strong interactions, where soft pions
interact with a massive hadron, or gluons  interact with 
a heavy quark.  This is reviewed in ref. \cite{hpet}. I
will simply take over the methodology to the case at hand.

The first observation is that since $m_H \gg \mu$, the
$4$-velocity of $H$, $v_{\mu}$, is approximately conserved in
collisions with soft gravitons, to within ${\cal O}(\mu/m_H)$. Thus the
$H$ momenta have the form,
\begin{equation}
p_{\mu} = m_H v_{\mu} + k_{\mu}; ~~~~ k_{\mu} \sim {\cal O}(\mu).
\end{equation}
For the simple case considered here, $v_{\mu}$ is common to all the $H$
particles involved, since their relative momenta were assumed to be
of order $\mu$.
We perform a field redefinition of the scalar field to remove the large
{\it fixed} component of the momentum, $m_H v$, 
\begin{equation}
H_v(x) \equiv \sqrt{2 m_H} e^{-i m_H v_{\mu} x^{\mu}} H(x).
\end{equation}
The $H_v$ field thereby has residual momentum $k_{\mu} \sim {\cal
  O}(\mu)$, just like the gravitons. Because $v_{\mu}$ is a Minkowski
space vector, not a generally covariant vector, it is important to note
that the generally covariant derivative for $H_v$ is $i m_H v_{\mu} +
\partial_{\mu}$, rather than just $\partial_{\mu}$ for $H$.

The general form of the $\mu$-scale effective lagrangian in this sector is,
\begin{eqnarray}
{\cal L}_{eff}(\mu) &=& \sqrt{-g} \{ \lambda + 2 M_{Pl}^2 R + 
\frac{1}{2 m_H} g^{\mu \nu} (-i m_H v_{\mu} + \partial_{\mu}) \overline{H}_v
 (i m_H v_{\nu} + \partial_{\nu}) H_v \nonumber \\
&-& \frac{m_H}{2} \overline{H}_v H_v +
 ... \} ~.
\end{eqnarray}
 The effective lagrangian is manifestly
generally coordinate invariant, a symmetry of the starting point,
eq. (5).
The ellipsis contains higher dimension operators constructed from
$g_{\mu \nu}$ and $H_v$, including local $H$ self-interactions,
 whose effects are small at large distances (but can be systematically
 included).
If $\lambda = 0$, we can expand the quantum theory about a Minkowski
vacuum, $g_{\mu \nu} = \eta_{\mu \nu}, H_v = 0$. 
In the  frame of
the $H$ particles, given by $v = (1, \vec{0})$, eq. (10) then becomes,
\begin{eqnarray}
{\cal L}_{eff}(\mu) &=& \sqrt{-g} \{2M_{Pl}^2 R + 
\overline{H}_v (i \partial_0 +
\frac{\partial^2}{2 m_H}) H_v - \frac{m_H}{2 M_{Pl}} h^{00} 
  \overline{H}_v H_v \nonumber \\
&+& {\cal O}(\frac{\mu}{M_{Pl}}) + ... \},
\end{eqnarray}
 describing 
non-relativistic $H_v$ particles coupled to gravity, predominantly
 through their gravitational ``charge'', $m_H/M_{Pl}$. 
 After gauge-fixing the
gravitational fields (see for example ref. \cite{veltman}), we
can integrate out $1$-graviton exchange  
which dominates the $H_v$
interactions, to obtain a non-local Newtonian potential interaction
between $H_v$ particles. If $\lambda \neq 0$, the field theory must be 
expanded about an ``expanding universe'' metric rather than Minkowski space.


Since eq. (10) describes a quantum field theory, we may ask if
$\lambda$ can  naturally be as small as eq. (4) 
under quantum corrections within this effective theory. The answer is
yes! Conceptually, {\it all} the
fields in eq. (10) have their momenta cut off at $\mu$, because of the
field redefinition, eq. (9). Thus although 
 by power-counting we estimate that $\lambda$
should get radiative corrections of order four powers of the cutoff,
this is just $\mu^4$, which for $1/\mu \sim 1$ mm, is well within the
experimental bound, eq. (4). In fact if we adopt dimensional
regularization, the $H_v$-loop corrections to $\lambda$ vanish.

The  effective  field theory described above
 reproduces some familiar phenomena of classical general
relativity, such as the Newtonian force between
non-relativistic masses, and gravitational radiation.
 On the other hand the effective
 theory is fully quantum mechanical and unitary in its domain of
 validity, and predicts inherently quantum corrections to the classical
 approximation. (An example  of such corrections is  described in
 ref. \cite{donoghue}.) Yet, it has a naturally small cosmological constant.
For these reasons it is  a useful conceptual link between microscopic physics
 and classical general relativity.

\subsection{The Cosmological Constant Problem}

The $\mu$-scale effective theory given by eq. (10) and the classical
effective theory of eq. (1) both share the same cosmological
constant, $\lambda$. I will focus on eq. (10) since it is a
straightforward quantum effective field theory, though analogous
statements follow for classical general relativity, eq. (1).
If we do not look beyond the effective theory of eq. (10),
$\lambda$ can naturally satisfy the experimental bound, eq. (4), as
pointed out above.
However, our present point of view is that $\lambda$ is
determined by {\it matching} the effective theory of eq. (10) with 
 the more fundamental theory of eq. (5), or
conceptually, by integrating out the physics below $v$. 
If one fixes a particular regularization and renormalization scheme, say
dimensional regularization with minimal subtraction, one can actually
perform the matching computations. Here we only require the results of 
simple power-counting, which  shows that
 $M_{Pl}$ is negligibly renormalized in matching at $\mu$, 
while by contrast, $\lambda$ is quartically sensitive to the
 mass scales of the SM, so that,
\begin{equation}
\lambda = \lambda_0 + {\cal O}(v^4).
\end{equation}
 We have no way
of understanding why the physics above the weak scale, which determines
$\lambda_0$, should so precisely cancel against the ${\cal O}(v^4)$ SM
contributions, in order for eq. (4) to hold.

Thus I conclude, although eq. (5) reduces to the SM at short distances,
(eq. (7)), reproducing all accelerator experiments, and though it reduces
to eqs. (1) and (10) at macroscopic distances, thereby accomodating all
gravitational measurements, {\it it is not 
  the correct effective theory below the weak scale} because it
involves a fantastic and inexplicable fine-tuning of $\lambda$. We must
therefore see what room we have for changing the weak scale theory
without destroying its highly non-trivial theoretical consistency and
agreement with experiment.

What seems highly significant to me is this. The cosmological constant
is usefully thought of as a non-derivative graviton self-coupling 
(which de-stabilizes Minkowski spacetime). Quantum corrections to
$\lambda$ come from loops of massive SM states, coupled to
external graviton lines at essentially zero
momentum. Therefore necessarily, 
these massive SM states are {\it far off-shell}. On the other hand,
experimentally we have only tested the gravitational couplings of SM
states which are {\it very nearly on-shell}.\footnote{By contrast note
 that accelerator
experiments {\it have} very successfully probed highly  virtual, purely
 SM effects, in the form of running couplings and precision electroweak 
tests.} For example, the $H$
particles of the previous subsection are very  nearly on-shell in the
domain of validity of the $\mu$-scale effective theory. It follows that
 all  the large quantum corrections to $\lambda$ from the  weak-scale
 theory of eq. (5) come from a tremendous theoretical extrapolation to
 the regime where gravitons couple to SM particles which are far
 off-shell.  We can
hold out some hope that the CCP can be avoided by a different
weak scale effective theory, which however still reduces to 
 eqs. (1) and (10) in the domain of on-shell SM matter coupled to
soft gravitons. 

\subsection{Constraints on alternative weak-scale theories}

In thinking about alternative effective theories, it is crucial to
observe two powerful fundamental principles, at least as far as physics
below the weak scale is concerned.
First,  to quite large distances, spacetime appears as a Minkowski
continuum. It also appears to be true down to distances of order $1/v$,
 since the highly successful SM 
loop computations depend sensitively on this assumption. 
Secondly, nature is quantum
mechanical, at least up to weak scale energies. Furthermore, it is
difficult to perturb the quantum principle withut leading to physical
absurdities.
 Therefore it would appear that we cannot
seriously doubt the principles of (local) special relativity and quantum
mechanics in the gravitational sector below the weak scale. These two
principles impose very severe constraints on model-building.
Taken with the experimental success of general
relativity at large distances they {\it necessarily} imply the existence
of a massless spin-two particle, the graviton, which must underlie any
effective theory of gravity. Furthermore, this effective theory 
 {\it must} obey the gauge symmetry of
general coordinate invariance \cite{spin2}. 
 This is similar to the case
of light  spin-one particles, where a gauge symmetry is needed to
decouple unphysical degrees of freedom, but in the case of spin-two the
gauge symmetry is unique!\footnote{There have been suggestions that
  general coordinate invariance can be replaced by {\it restricted}
invariance under coordinate transformations with unit Jacobian. However,
both classically and quantum mechanically this is
precisely equivalent to a generally invariant theory with an arbitrary
(but not naturally small) cosmological constant. See ref. \cite{review}
for a brief review, plus references.} 

Now, if we restrict ourselves to the  minimal particle content, namely
the SM particles and the graviton, the form of the effective
theory is given by  eq. (5), this being the most general invariant form
that reduces to the SM when
$1/M_{Pl} \rightarrow 0$, and containing the kinetic term for the
graviton field. But eq. (5) is just the effective theory we are trying to
avoid. Thus we conclude that new particles associated with gravity must
be present. They must be very light indeed in order to remain in the
effective theory down to the  very low energies necessary to 
 protect the cosmological constant, as has recently been
emphasized in ref. \cite{beane}.

Unfortunately, all proposals to couple extra particles to eq. (5) have
failed to cure the naturalness problem. Generic addition of
extra light particles does not evade the simple power-counting which says
that the cosmological constant is quartically sensitive to the {\it highest}
mass scales in the theory. Supersymmetrizing eq. (5) does in fact
stabilize a suitably small $\lambda$. However this requires
supersymmetry to be unbroken in the SM sector to  very high
precision, in order to suitably 
reduce the ${\cal O}(v^4)$ contribution in
eq. (12). Experimentally however, we know that supersymmetry is badly broken in
this sector. Other than supersymmetry the only other special symmetry
that can control the cosmological constant is conformal symmetry. This
is also badly broken in nature, but there have been several attempts to
make $\lambda$ a dynamical field that relaxes to zero as a consequence
of the conformal anomaly, similarly to the way an axion can relax the
strong interactions to a CP-conserving vacuum  in the presence of a
$\theta$ angle, as a consequence of the
axial anomaly. For the CCP 
all such attempts have failed for the general reason
described in ref. \cite{review}. 
To summarize, while we can always
weakly couple eq. (5) to new light particles, there is no reason for these to
significantly reduce the ${\cal O}(v^4)$ SM loop contributions to
$\lambda$. 

The seemingly impossible predicament posed by the CCP has given rise to 
proposals which play by different rules from those we have adopted. They cannot
be evaluated within any local
effective theory and  are 
difficult to test experimentally. It is possible that one of these
proposals is nevertheless true. 
Perhaps the best-known 
is Coleman's wormhole proposal \cite{coleman} \cite{review}. Here,
wormhole physics, just below the Planck scale, gives rise to peculiar
non-local effects (from the viewpoint of our macroscopic spacetime),
whereby the fundamental ``constants'' of nature become dynamical
variables, {\it but
  without any local spacetime variations.} The relevant path integral
 is infinitely peaked at values of these
constants such that the bottom-line
 cosmological constant vanishes, $\lambda = 0$.

The present paper describes a deliberately restricted  search for a
resolution of the CCP which can be described by a natural
effective theory, expressed in terms of
{\it local} degrees of
freedom. This is the time-honored approach taken
towards other naturalness problems such as the the Strong CP problem or
the Higgs naturalness problem. However, the arguments of
this section seem to  suggest that we are at an impasse. There may be a
way out though, as suggested by the following parable.

\section{An Analogy}

In this section I describe a naturalness problem, analogous to
the CCP, which occurs within a toy model
universe. This toy problem has the advantage of involving only the
familiar quantum field theory of particles with spins less than or equal
to one. Nevertheless, the resolution sheds light on how the CCP 
might be resolved. The model consists of two sectors, a
toy ``Standard Model'' (TSM) accounting for short-distance
``laboratory'' physics, and a toy ``gravity'' (TG)  only
noticeable at very large distances. 

The TSM is simply the quantum electrodynamics of eight identical flavors
of charged fermions, $\psi_a$, $a = 1,..., 8$. I will cut off the
electromagnetic interactions at larger than laboratory distances by
giving the photon a very small mass.\footnote{Of course, in the real
  world the electromagnetic force is negligible on large distance scales
  because
  of the neutrality of massive gravitating  objects, like planets and stars. 
The toy photon mass makes for a simpler story.} Recall that for an {\it
abelian} gauge field, a mass term is both renormalizable and naturally
small (only receiving logarithmic quantum corrections). The
renormalizeable and natural TSM theory is then given by
\begin{equation}
{\cal L}_{TSM} = \overline{\psi}_a (i \slashchar{D} - m) \psi_a -
\frac{1}{4} F_{\mu \nu}^2 - \frac{m_A^2}{2} A_{\mu}^2, 
\end{equation}
with $m_A \ll m, ~\alpha_{em}(m) \ll 1$.
We will consider the TSM to have been tested at lab momenta, very
roughly of order $m$ (where the photon mass is negligible), and to a
precision given by $\epsilon \ll \alpha_{em} \ll 1$. For example,
  our momentum resolution is of order $\epsilon m$, and we are
insensitive to $n$-loop QED effects for $n$ such that $\alpha_{em}^n <
\epsilon$. Nevertheless I will consider $\epsilon$ to be small enough
that eq. (13) has been non-trivially tested as a quantum field theory.

On the other hand, TG corresponds to  the observation of a very weak
classical scalar Yukawa force, $y^2 e^{- m_{\phi} r}/r^2$,
between non-relativistic $\psi$ particles\footnote{Unlike the real world,
  in the toy universe the photon does not ``gravitate''.} 
over very large distances and
times, 
\begin{equation}
r, t \gg \frac{1}{m_{\phi}} \gg \frac{1}{m_A}. 
\end{equation}
Notice this implies that the exponential suppression is always turned on
in the Yukawa force, but clearly it is still the dominant force at very
large distances. 
The TG force is too weak to be observed at short distances in the lab,
against the background of electromagnetism,
but is seen outside the photon range. To be concrete let us take,
\begin{equation}
y^2 \sim \epsilon.
\end{equation}
 The mass scale $m_{\phi}$ is
extremely small,
\begin{equation}
m_{\phi} \ll m e^{-1/\epsilon}.
\end{equation}
At the purely classical level this is  acceptable, as is a very small or
zero cosmological constant in  classical general relativity. 

The minimal relativistic quantum field theory incorporating both the TSM
and TG necessarily associates a scalar field, $\phi$, with the Yukawa force,
\begin{eqnarray}
{\cal L} =  \overline{\psi}_a (i \slashchar{D} - m + y \phi) \psi_a -
\frac{1}{4} F_{\mu \nu}^2 - \frac{m_A^2}{2} A_{\mu}^2 
+\frac{1}{2} (\partial_{\mu} \phi)^2 - \frac{m_{\phi}^2}{2} \phi^2 -
g_{\phi} \phi^4,
\end{eqnarray}
where the scalar coupling $g_{\phi}$ is included for renormalizability,
though it is too small to observe and plays no further role. Eq. (17) is
the analog of eq. (5). Like eq. (5), it suffers from a naturalness
problem. Here, the problem is why the scalar mass, $m_{\phi}$,  is so
small, despite much larger quantum corrections coming from TSM
$\psi$ loops. Standard power-counting and  eqs. (15, 16) give,
\begin{equation} 
\delta m_{\phi}^2 \sim y^2 m^2 \gg m_{\phi}^2.
\end{equation}

Physicists of the toy universe may note that a small scalar mass is
stabilized by supersymmetry. But the fact that no $\psi$ superpartners
have been observed for energies well above $m$ means that
supersymmetry is badly broken, and eq. (18) still holds. This is closely
analogous to the situation with the cosmological constant and
supersymmetry in the real world.

Other than supersymmetry there is no mechanism by which a
 weakly coupled fundamental scalar can
naturally avoid corrections like eq. (18).
One might think that the spin-$0$ particle 
 could be fundamental and light if it
is a Nambu-Goldstone boson of a spontaneously broken symmetry, but this
possibility can be ruled out as well. Even though fundamental
Nambu-Goldstone bosons are naturally massless, their ``decay constants''
are naturally of order the highest scale in the theory, in the present
case, $f_{\pi} \sim m$. The fact that the spin-$0$ particle has 
non-derivative
couplings to the $\psi$'s means that the spontaneously broken symmetry
must also be explicitly broken. The same explicit breaking which gives
rise to a Yukawa coupling, $y$, naturally gives rise to a
pseudo-Nambu-Goldstone boson mass-squared of order
 $y f_{\pi}^2 \sim y m^2$, which is incompatible with eq. (16).   

The only remaining means of obtaining a naturally light spin-$0$
particle is to make it a composite, like a hadron, {\it with a very low
compositeness scale.} Even this approach offers no comfort at first
sight. The basic reason is that the 
$\phi$ self-energy estimate due to TSM loops,
 yielding eq. (18), is performed at
essentially zero external momentum, and so is completely insensitive to
whether $\phi$ is composite or fundamental. Thus in any model where
$\phi$ has a low enough compositeness scale to naturally satisfy eq. (16), 
it will be impossible to arrange for a Yukawa coupling as large as in
eq. (15). I will illustrate this with a
specific example. Suppose we try to make $\phi$ a scalar glueball of a
Yang-Mills sector, with a confinement scale $\Lambda_{YM} \ll m
e^{- 1/\epsilon}$, which sets the glueball mass. To obtain a
Yukawa coupling to the $\psi$'s we can use a higher dimension
interaction,
\begin{equation}
\delta {\cal L} = h \frac{\overline{\psi} \psi~ tr ~G_{\mu \nu} G^{\mu
    \nu}}{m^3},
\end{equation}
where $h$ is a dimensionless coupling, and tr$G_{\mu \nu} G^{\mu \nu}$ 
is the Yang-Mills operator that interpolates the glueball, $\phi$,
\begin{equation}
<0|  tr G_{\mu \nu} G^{\mu  \nu} |\phi> ~ \sim ~\Lambda_{YM}^3.
\end{equation}
Recall that a non-renormalizable interaction such as eq. (19) is
acceptable within  effective field theory. Taking the effective
theory cutoff to be of order $m$, the energy scale probed in the lab, the
theory remains weakly coupled at the cutoff provided $h \ll 1$. (In fact
we must have $h < \epsilon$ in order for the $\psi$-gluon interactions to
have not been directly seen in the lab.) Therefore we arrive at the 
unsatisfactory result,
\begin{equation}
y \sim h \frac{\Lambda_{YM}^3}{m^3} \ll e^{- 3/\epsilon},
\end{equation}
in contradiction to eq. (15). 

I hope to have convinced the reader that, like the CCP,
 this toy naturalness problem seems to leave no room for
manoeuvre. However, this is a false impression.

Fortunately, compositeness {\it does} allow the resolution of the
naturalness problem. In order to invalidate the reasoning behind the
large quantum corrections to $m_{\phi}$ from the TSM, we must not only
take $\phi$ to be a composite light hadron, but we must also
consider the massive particles it interacts with in TG to be heavy
hadrons containing the TSM $\psi$ particles as heavy quarks!  The
specific resolution I have in mind is given by,
\begin{eqnarray}
{\cal L} &=& \overline{\psi}(i \slashchar{D} - m) \psi +
 \overline{u}(i \slashchar{D} - m_u) u + \overline{d}(i \slashchar{D} - m_d) d 
\nonumber \\
&-& \frac{1}{4} F_{\mu\nu}^2 - \frac{m_A^2}{2} A_{\mu}^2 - \frac{1}{4} tr
G_{\mu \nu}^2 +\frac{\theta \alpha_s}{8 \pi} tr ~ \tilde{G}_{\mu \nu} G^{\mu
  \nu},
\end{eqnarray}
where I have introduced an $SU(3)_{QCD}$ gauge theory, under which the
$u$ and $d$ quarks are triplets, and the eight $\psi_a$ TSM fields form
an $SU(3)_{QCD}$ adjoint representation. Only the $\psi$'s are
electrically charged however. 

The $\phi$ is a particular combination of  the three QCD pions,
$\vec{\pi}$, made out of $u$
and $d$ quarks, while
the heavy fermion it interacts with hadronically in TG, $\Psi$, is a
composite of the adjoint $\psi$ quark and glue. The light quark masses
$m_{u,d} \ll \Lambda_{QCD}$ are needed to produce non-derivative Yukawa
couplings of $\phi$ to $\Psi$, and to generate a small $m_{\phi}$. The
Yukawa coupling we need, $\phi \overline{\Psi} \Psi$, breaks isospin
symmetry under which the pions form a triplet, whereas $\Psi$ is a
singlet. The requisite isospin breaking is arranged by taking $m_u \neq
m_d$. Another  technicality is that QCD is normally a parity-conserving
theory, so a single pseudo-scalar pion cannot couple to the scalar
$\overline{\Psi} \Psi$ as required. I have therefore added an order one
CP-violating $\theta$-term.\footnote{In fact, even for $\theta \neq 0$,
the Yukawa coupling is not generated at first order in $m_{u,d}$
because of a vacuum re-alignment induced by
$\theta$. However, the Yukawa coupling {\it is} generated at higher
order in $m_{u,d}$.}
The resulting Yukawa coupling is then of order 
a small power of $m_{u,d}/\Lambda_{QCD}$, while
\begin{equation}
m_{\phi}^2 \sim (m_u + m_d) \Lambda_{QCD}. 
\end{equation}
By taking $\Lambda_{QCD} \sim m e^{- 1/\epsilon}$, we can
consistently choose $m_{u,d} \ll \Lambda_{QCD}$ so that $y \sim
\epsilon$ and $m_{\phi} \ll  m e^{- 1/\epsilon}$, as desired!

There are three issues we would like to understand better: (i) Why is
the composite QCD structure not already observed in the long distance TG
sector? (ii) Why is the composite structure not visible in the lab?
(iii) How does the composite structure cure the $\phi$ mass of extreme
sensitivity to the TSM mass scale, $m$? Some of the discussion will be
 similar to
that of ref. \cite{okun}, where a model with very small $\Lambda_{QCD}$
was also considered.

(i) The interactions of very 
low-energy pions with slow heavy $\Psi$ hadrons
can be described using a Heavy Hadron Effective Lagrangian (reviewed in
ref. \cite{hpet}), 
\begin{equation}
{\cal L}_{eff} =  \overline{\Psi} (i \partial_0 + \frac{\partial^2}{2 m} + 
y \phi ) \Psi + \frac{1}{2} (\partial_{\mu} \phi)^2 -
\frac{m_{\phi}^2}{2} \phi^2 + ... ~.
\end{equation}
Recall that $\phi$ is some linear combination of the $\vec{\pi}$ fields
depending on $m_{u,d}$ and $\theta$.
The ellipsis contains terms  whose effects on $\Psi$ are negligible
at the very low momentum transfers, $p \ll m_{\phi} \ll 
\Lambda_{QCD}$, corresponding to eq. (14). These include the 
 higher dimension 
couplings (suppressed by powers of $\Lambda_{QCD}$) of $\phi$ to itself
and to the $\Psi$, and all couplings involving $\vec{\pi}$ fields other 
than $\phi$. Eq. (24) is just the analog of eq. (11). Integrating out
$\phi$-exchange, which dominates the $\Psi$ interactions at long
distance,   yields the simple 
Yukawa potential. The next lightest state above
 the $\phi$
that can be exchanged between $\Psi$'s is a two-pion state. But in the
regime of eq. (14), even the two-pion exchanges are exponentially
suppressed relative to single-$\phi$ exchange.

 As far as eq. (24) is concerned the scalar mass is
naturally small because of the very low cutoff on the effective theory.
Compositeness effects are invisible because the
compositeness scale is too high compared with the (virtual) 
$\phi$ momenta corresponding to eq. (14).
 We see that the first-conjectured form of the
TG field theory is wrong. $\phi$ does not interact with the $\psi_a$
quarks, but rather with the hadronic ``brown muck'' of the heavy $\Psi$
hadron.

One might worry that there can be excited $\Psi$
composites, which have different Yukawa couplings $y$, but over the time
scales of eq. (14)
such states would decay to the lowest stable $\Psi$ state. 
A minor technical dynamical assumption that must be made (but which
fortunately has no analog in the real CCP) is that any exotic
composites of $\psi$ and two or more light quarks are heavy enough
 to decay to $\Psi$ via pion emission, so that their possibly
different Yukawa couplings are not seen.
 
(ii) Typical lab momenta are of order $m$, where the running QCD
coupling is weak. In the limit where it vanishes, the TSM sector
completely decouples from the QCD sector. We can work out the actual
value of the coupling renormalized at the laboratory momentum
resolution, $\alpha_s(\epsilon m)$, using the one-loop QCD 
$\beta$-function and the fact that we have already chosen 
$\Lambda_{QCD} \sim m e^{- 1/\epsilon}$. The result is, 
\begin{equation}
\alpha_s(\epsilon m) \sim \epsilon \ll \alpha_{em}.
\end{equation}
Therefore QCD-induced $\psi$ momentum transfers larger than $\epsilon m$
have amplitudes suppressed by $\epsilon$, so they are too small to be
seen against
TSM interactions. On the other hand, amplitudes where the QCD-induced
momentum transfers are less than $\epsilon m$ remain unsuppressed,
corresponding to soft 
radiation of light hadrons and excitation
of the $\Psi$ resonances. But such momentum transfers are smaller than
our momentum resolution.  
The QCD sector is therefore invisible in the lab!
 Note, it is only the $\psi$ particles which  feel the
electromagnetic force and determine the outcome of lab experiments. 
 
In the absence of the electromagnetic interaction the $\psi$'s would
also form heavy quarkonium bound states, but with electromagnetism the
QCD interactions will only negligibly perturb the electromagnetically 
 bound states.

(iii) We now see that the unnaturally large quantum corrections in
eq. (18) arose because
of  $\Psi$ loops, where the $\Psi$ appears far off-shell, with 
 a  Yukawa coupling to  $\phi$. But this simple coupling is only
 valid in eq. (24), where the  $\Psi$ is nearly
on-shell. The extrapolation off-shell is completely invalid since the 
$\Psi$ compositeness
scale is very low. The true quantum corrections to the
$\phi$ mass from the TSM sector require knowledge of the full QCD
dynamics. We will now correctly compute the $m_{\phi}$ sensitivity to
$m$. To make the question precise let us fix some ultraviolet cutoff,
$\Lambda_0 \gg m$, relative to which we can measure masses. This could be
the scale of some new physics beyond the toy standard model. We also fix
$\alpha_s(\Lambda_0)$ and ask how $m_{\phi}/\Lambda_0$ changes as a
function of $m/\Lambda_0$. We already have the mass formula for
pseudo-Nambu-Goldstone bosons, eq. (23). We can integrate out the effects
of the very heavy (adjoint) quark because of the asymptotic freedom of
QCD. The dominant behavior follows from the one-loop renormalization
group. The infrared renormalized quark
mass parameters that appear in eq. (23) are the result of running down from
$\Lambda_0$. To one-loop order however, this mass renormalization 
 is independent of the
heavy quark mass, $m$. Only $\Lambda_{QCD}$ is changed at one loop because
the heavy quark slows down the running of $\alpha_s$ between $\Lambda_0$
and $m$. A standard perturbative matching computation  then leads to, 
\begin{equation}
m_{\phi}^2/\Lambda_0^2  \sim (m/\Lambda_0)^{\frac{24}{29}}
 e^{- \frac{12 \pi}{29 ~\alpha_s(\Lambda_0)}} 
~(\frac{m_u + m_d}{\Lambda_0}).
\end{equation}
We see that $m_{\phi}$ is not unnaturally sensitive to $m$. 
Doubling $m$  only leads to a doubling of $m_{\phi}^2$, as compared to
the extreme and unnatural sensitivity implied by the naive result,
eq. (18). The quadratic sensitivity of scalar radiative corrections  
to the ultraviolet scale $m$ has been eliminated by having no 
scalar degree of freedom present at $m$, only quarks and gluons. These
constituents of the scalar are only logarithmically sensitive to $m$.

To summarize, we were able to resolve the toy naturalness problem by
giving the TG sector a very low compositeness scale and making
the ``gravitating'' TSM particles into constituents of
composite states. The toy
composite ``graviton'' $\phi$ interacts with the compositeness 
``halo'' that
surrounds the TSM particles. At very low momenta this is
indistinguishable from a direct coupling to the nearly on-shell TSM 
particles. 
If this is extrapolated to when the TSM particles are far off-shell, one
runs into the naturalness problem. In reality though, the off-shell
contributions are cut off by compositeness. The naive extrapolation
misses this composite softness of the interactions in the TG
dynamics. On the other hand the compositeness interactions are
invisible in the lab, compared with the much stronger hard interactions 
of the
TSM. The obvious regime to discover the compositeness dynamics is at
intermediate distances, where TSM interactions are still neutralized but
compositeness effects are unsuppressed in the TG dynamics.

\section{The Effective Particle-String Scenario}

The moral of the previous section is that the power-counting
that points to the inevitability of the CCP only holds if the graviton
is fundamental, not if it is ``composite''. To exploit this observation 
 we must
ensure that {\it there simply is no graviton}  at the energies at which 
we integrate out SM particles, en route to
obtaining the long-distance theory of gravity. At these SM energies
 there should  only be the 
degrees of freedom which will bind into the graviton at much 
lower energies.   A second requirement is that 
  the SM particles must couple to composite gravity, and yet their
  couplings to other SM particles must be point-like at least down to
  $1/v$ distances. Unlike the case of
the scalar $\phi$ in the toy model though, 
the compositeness of the graviton
cannot be accomplished within the ordinary Minkowski space 
quantum field theory of point
particles. This is due to the following very general theorem 
 \cite{theorem}: a theory in Minkowski space which
admits a well-behaved, conserved energy-momentum tensor cannot have a
graviton in its spectrum. 

Fortunately, string theory evades this theorem and gives a sensible
meaning to graviton compositeness. Though formulated in Minkowski space,
its energy-momentum tensor is not ``well-behaved'' and there is a
massless graviton in the spectrum, as discussed in
 ref. \cite{discrete}. In terms of the well-known similarity
between string theory and QCD  (which of course was
important historically for the discovery of string theory \cite{gsw}),
the graviton can be thought of as a massless ``glueball'' of string
theory. Now in QCD there are sum rules that can be derived in terms
of the fundamental description which look miraculous or finely-tuned in
terms of the hadronic description. They are not enforced by any symmetry
but by the special nature of the dynamics. The same is true in string
theory with respect to the cosmological constant.
  
For simplicity let us
consider the case of the perturbative bosonic string 
in 26 dimensions \cite{polchinski}. 
There are effectively two parameters, the string mass-scale, $m_{st}$,
which plays the role of the graviton compositeness scale,
and the string coupling,
$g_{st}$. For $g_{st} \ll 1$, the string spectrum  corresponds to an
infinite number of ``composite'' particle-modes of varying spins and 
masses, including a graviton. The Planck scale is very large, $M_{Pl} \gg
m_{st}$. 
If we ignore the string principle, we can compute
the $1$-loop contributions to the cosmological constant from each of the
particles. Each contribution is quartically
sensitive to the particle mass, and there are an infinite number of such
contributions. Clearly we must introduce an ultraviolet cutoff, which
cuts off both the infinity of contributions and the infinity in each
contribution. This still leaves many large
contributions to the cosmological constant. 
If we want the renormalized
cosmological constant to come out very small, we must also add a
 counterterm chosen very precisely to finely cancel the large
 $1$-loop contributions. Of course, in string theory the sum of one-loop
 diagrams plus counterterm is not calculated in this piecemeal fashion, 
but rather at one stroke. The result is an ultraviolet
 finite cosmological constant,
$\lambda \sim m_{st}^{26}$. Note, this is just the power-counting
dependence on an ultraviolet cutoff, $m_{st}$,  in 26-dimensional general
relativity. This illustrates our  expectation that the
 compositeness scale should cut off the divergences of general relativity.
Since the $26$-dimensional Planck scale is
given by 
\begin{equation}
M_{Pl} = \frac{m_{st}}{g_{st}^{1/12}},
\end{equation}
$\lambda$ can be made arbitrarily small compared to $M_{Pl}^{26}$ 
 by taking small
enough $g_{st}$. In Planck units this corresponds to a very low tension 
string theory. 

However, in the usual string formulation of particle physics, the SM
particles are also identified as string vibrational modes, and we must have
$m_{st} > v$ so that the stringy excitations of the SM particles
are too massive to appear in present-day experiments. Strings with such
a large string-scale cannot solve the CCP. In fact we can integrate out the
excited string states and return to eq. (5) and its unpleasant 
consequences. Instead, we wish to
pursue the possibility that there are strings in the gravitational sector 
with extremely low string-scale, 
$m_{st} \ll v$, but the SM particles are not themselves
made of {\it these} strings. Instead SM particles are point-like, at
least up to the weak scale. 
Just as the heavy quarks of the last section were surrounded by a light
hadronic halo to form a heavy meson,  the SM particles may be
surrounded by a stringy halo with which the graviton string mode
interacts. 

As yet there are no fully realisitic candidates known within string
theory, which are point-like on the string scale and can be identified
with the SM particles. However the recently discovered solitonic
D-branes \cite{dbranes} do possess some promising qualitative features.
For example,
0-branes are point-like objects with masses much larger than $m_{st}$,
which can probe a continuum spacetime down to distances much shorter
than $1/m_{st}$ \cite{douglas}. At long distance their interactions
conform to general relativistic expectations in terms of graviton
exchange. At distances smaller than $1/m_{st}$ the composite graviton is
an entirely inappropriate degree of freedom, and the force between
0-branes becomes intrinisically stringy \cite{douglas}.


I therefore propose that in nature the SM particles are dynamically 
more akin to 0-branes than they are to perturbative string modes such as
the graviton. Since string theories are only consistently formulated with
supersymmetry, it remains a problem to explain how supersymmetry ends up
badly broken in the SM sector. Nevertheless, supposing this is possible,
we would like to explicitly understand how the cosmological constant can be
cut off by the scale of graviton compositeness, $m_{st}$, rather than
being sensitive to the much larger SM masses. Below I offer a picture of
how this might work. I can make no pretence of rigor.

\subsection{How the particle-string might solve the CCP}

Let us consider a simple,  abstracted version of our problem. To
eliminate the complication of supersymmetry-breaking and
compactification,  let us simply work
within bosonic string theory in 26 euclidean dimensions (turning a blind
eye to the existence of a tachyon). This will be our gravitational
string sector. Let the 0-brane of this theory represent a ``SM
particle'', with mass $m \sim m_{st}/g_{st}$
\cite{dbranes}. For $g_{st} \ll 1$, the 0-branes are much more massive
than the string scale.
Strings are permitted to end on the 0-brane worldline,
the attached string constituting a string ``halo''.  Closed
 strings, including the graviton, are emitted and absorbed by 
this halo, inducing gravitational interactions for the
0-branes. We want to estimate the contributions of 
virtual 0-brane loops to the cosmological
constant, $\lambda$.\footnote{Strictly speaking,  the notion of a 0-brane
perturbative loop expansion is ill-defined, since these 0-branes are so 
massive that their gravitational couplings are large. I will however use
this language since it is the most familiar one, and because it is
likely to apply to a more realistic construction.}  
The naive power-counting guess, ignoring the string
principle, would be  $\lambda \sim m^{26}$.
 I will argue that the cosmological constant is instead set by the
 graviton compositeness scale, $m_{st}$, so that $\lambda \sim m_{st}^{26}$. 

The simplifying consideration is that the euclidean 
action for a particle of mass, $m$, will suppress 0-brane 
world-line loops which
are much bigger than $1/m$. Thus on the string scale 
 they are essentially point-like events
in spacetime, to which string worldsheet
boundaries  can attach.\footnote{This is very much like
ordinary quantum field theory, where integrating out a massive field
introduces local interactions for the light fields.} In the string
literature, such events are known as ``D-instantons''. The $1/m_{st}$
sized strings should be insensitive  to the tiny $1/m$-scale structure.

The contribution to the cosmological constant due to D-instantons has
been computed  and the result is finite and of order 
$e^{-1/g_{st}}$ \cite{d-instanton}. 
This result can be understood as follows. The 
cosmological constant correction
 is given by the sum of (first-quantized) connected string
diagrams with no vertex operators, where the string worldsheet 
boundaries attach to the D-instanton. The dominant contribution 
from a single worldsheet is of order  $1/g_{st}$, corresponding to
a disk topology, more complicated topologies being suppressed by powers
of $g_{st}$. The sign of this contribution requires a detailed
calculation and is negative. The dominant contribution from $k$
worldsheets is given by $k$ identical disks, whose boundaries attach to
the D-instanton. Their contribution is just the $k$th power of the
single-disk result, divided by a symmetry factor of $k$!. Summing over
$k$ gives the factor $e^{-1/g_{st}}$. 

Thus we expect that the contribution to the cosmological constant from
0-brane loops is suppressed by $e^{-1/g_{st}}$ (without being careful about
the prefactor) and is therefore negligible for $g_{st} \ll 1$!
Therefore the 
cosmological constant is dominated by the string-loop correction 
discussed earlier,  
\begin{equation}
\lambda \sim m^{26}_{st}.
\end{equation}

\subsection{Phenomenological aspects of this scenario}

A fully realistic effective particle-string theory 
has not yet been constructed. I will just
list some important features that it ought to have.

$\bullet$ The theory must contain SM particles and critical strings.
 The particles
must live in four spacetime dimensions and be point-like at least down
to $1/v$ distances. The string length scale and
compactification radii can however be much larger. Examples of
four-dimensional particle-like behavior co-existing with strings, and
large compactification radii ``seen'' only by the strings, have been
found and discussed in refs. \cite{strominger} \cite{shenker}.

$\bullet$ The theory must be unitary below the weak
scale. It is permitted
to break down above the weak scale, since we are not trying to guess the
very high energy physics.  

$\bullet$ The spin-2 graviton must be the only 
 massless non-SM  state with couplings to matter (unless they are
 even weaker than gravity). Then unitarity ensures that at long
 distances the dynamics reduces to general relativity \cite{spin2} 
\cite{gsw}. For distances of order $1/m_{st}$ or smaller, the massive
string physics will become important and general relativity must break
down. The fact that gravity has already been
tested at distances of a few centimeters without deviation from Newton's
Law, gives the bound,
\begin{equation}
m_{st} > 10^{-5} {\rm eV}.
\end{equation}

$\bullet$ The compositeness of gravity must make the cosmological
constant insensitive to the large SM masses, its size being set instead by 
the compositeness scale, $m_{st}$, 
\begin{equation}
\lambda \sim {\cal O}(m_{st}^4).
\end{equation}
This is also the power-counting result that follows from thinking of
$m_{st}$ as an ultraviolet cutoff for the effective theory of general
relativity. 

To satisfy the bound of eq. (4), we must have, 
\begin{equation}
m_{st} < 10^{-2} {\rm eV}.
\end{equation}
If the string compactification radii are of order $1/m_{st}$, the string
coupling is given by,
\begin{equation}
g_{st} = m_{st}/M_{Pl} = 10^{-30}.
\end{equation}
This may seem absurdly small, but recall that in string theory,  $g_{st} =
e^{- \langle {\rm dilaton} \rangle}$, and the stabilization of the
dilaton vev is still not understood. 
It may be related to the other absurdly small
number in nature, $v/M_{Pl} \sim 10^{-16}$. In any case, small $g_{st}$
is not technically unnatural.

$\bullet$ The new stringy physics must be negligible in SM
experiments. While the strings have typical length $1/m_{st} > 10^{-1}$ mm,
their couplings are so incredibly weak that they should not interefere with
the SM interactions. At lab momenta, the strings  should form an insubstantial
cloud about the SM particles. In particular, they should not upset the
theoretical agreement with SM experiments which are sensitive to very
small mass splittings, such as kaon mixing or atomic structure. This may
be of concern given eq. (29). 

$\bullet$ String theories are presently formulated with supersymmetry as
an essential ingredient for full consistency, 
yet supersymmetry must appear broken by at least $v$ in
the particle sector. This suggests a minimal supersymmetry breaking in
the string sector of order $v^2/M_{Pl}$. This scale may set the minimal
permissible string-scale. If so, $m_{st} > 10^{-4}$ eV.

~

Finally, let us consider how this scenario might be experimentally
tested. We expect that the nature of the gravitational
force should dramatically change for distances smaller than the
compositeness length scale, $1/m_{st}$, in a manner which cannot be
described by the exchange of a finite variety of massive particles (such
as the light scalars discussed in refs. \cite{beane} or \cite{moduli},
for example). For example, the interaction between a pair of 0-branes is
described at long distance in terms of the exchange of massless closed
string modes such as the graviton, while at short distance it is
described by open strings connecting the 0-branes \cite{dbranes}
\cite{douglas}. In this regime 
the gravitational force can become weaker with shorter distances! 

Eqs. (29) and (31) narrowly constrain the compositeness length scale at
which the radical
departures from general relativity (Newton's Law) must occur,
\begin{equation}
10 \mu {\rm m} < 1/m_{st} < 1 {\rm cm}.
\end{equation}
It is therefore our very good fortune that this is just the range over
which gravity will be sensitively tested in the experiment proposed in
ref. \cite{expt}. If composite gravity resolves the CCP as proposed
here, it will show itself in this experiment and be quite distinct from
any other ``fifth force'' phenomenon which can be described within field
theory!

\section{Conclusions}

The Cosmological Constant Problem was argued to be intractable as a
naturalness problem in effective field theory unless the graviton was
``composite'' with a scale of compositeness below $10^{-2}$ eV. The
standard model particles must also participate in this compositeness and
yet retain their point-like behavior in accelerator experiments up to
very high energies.  The only sensible version of
graviton compositeness that is known, is string theory. 
It was proposed that the
standard model particles inherit their gravitational interactions by
virtue of their
their stringy ``halos'', coupled to a gravitational string sector. The
string-scale plays the role of the compositeness scale. It was
argued that this stringiness can acceptably cut off  contributions
to the cosmological constant from ultraviolet mass scales. The mechanism is
reminiscent of the relative insensitivity of light hadron masses to heavy
quark masses in QCD. This was the basis for the detailed analogy
discussed in the paper. 

If this particle-string scenario
is realized in nature, it will lead to a dramatic breakdown of Newton's
Law on the millimeter scale, which  will be experimentally probed.
On the theoretical side, much work still remains in order to construct a
fully realistic effective particle-string theory and demonstrate its
requisite properties. The particle-string 
scenario considered here
would obviously also have deep implications for physics at the highest 
energy scales. 

Finally, 
it is worth keeping in mind  that there may be other, presently
undiscovered, 
manifestations of graviton compositeness that can also reduce the
sensitivity of the cosmological constant to ultraviolet mass
scales. Fortunately, independently of the form of graviton
compositeness which resolves the CCP, power-counting 
suggests that eq. (33) constrains the compositeness length
scale. Therefore, the composite behavior should still show up in upcoming
experimental tests of gravity at short distances.

\section*{Acknowledgments}
This research was supported by the U.S. Department of Energy under grant
\#DE-FG02-94ER40818. I wish to thank Tom Banks, Sekhar Chivukula, Andrew Cohen,
Nick Evans, 
Shamit Kachru, Martin Schmaltz and especially my
father, R. M. Sundrum, and my wife, Jamuna Sundrum, 
for useful conversations on the subject of this paper.

\end{document}